\documentclass[reprint,twocolumns,pre,amsmath,amssymb]{revtex4-1}
\usepackage{graphicx}
\usepackage{booktabs}
\usepackage{array}
\usepackage{hyperref}

\usepackage{bm}
\usepackage{times}

\usepackage{balance}
\usepackage{epic,eepic}
\usepackage{graphics}
\usepackage{epsfig}
\usepackage{subfigure}

\usepackage[latin1]{inputenc}
\usepackage{pstricks,pst-node,pst-text,pst-3d}
\usepackage{amsmath}
\usepackage{graphicx}
\usepackage{pst-grad}
\usepackage{pst-vue3d}
\newpsstyle{GradGrayWhite}{fillstyle=gradient,%
    gradbegin=blue,gradend=white,linewidth=0.1mm}%
\usepackage{graphicx}
\usepackage{pst-grad}
\usepackage{pst-vue3d}
\usepackage{soul}

\newcommand{\be}{\begin{equation}}
\newcommand{\ee}{\end{equation}}
\newcommand{\bea}{\begin{eqnarray}}
\newcommand{\eea}{\end{eqnarray}}

\usepackage{amsmath}
\usepackage{amsfonts}
\usepackage{amssymb}
\usepackage{bm}
\usepackage{color}
\usepackage{graphicx}

\begin{document}

\title{Avalanche statistics during coarsening dynamics}

\normalsize

\author{F. Pelusi} \affiliation{Department of Physics and INFN, University of Tor Vergata, \\ Via della Ricerca Scientifica 1, 00133 Rome, Italy} 
\author{M. Sbragaglia} \affiliation{Department of Physics and INFN, University of Tor Vergata, \\ Via della Ricerca Scientifica 1, 00133 Rome, Italy}
\author{R. Benzi} \affiliation{Department of Physics and INFN, University of Tor Vergata, \\ Via della Ricerca Scientifica 1, 00133 Rome, Italy} 

\begin{abstract}
We study the coarsening dynamics of a two dimensional system via lattice Boltzmann numerical simulations. The system under consideration is a biphasic system consisting of domains of a dispersed phase closely packed together in a continuous phase and separated by thin interfaces. Such system is elastic and typically out of equilibrium. The equilibrium state is attained via the coarsening dynamics, wherein the dispersed phase slowly diffuses through the interfaces, causing domains to change in size and eventually rearrange abruptly. The effect of rearrangements is propagated throughout the system via the intrinsic elastic interactions and may cause rearrangements elsewhere, resulting in intermittent bursts of activity and avalanche behaviour. Here we aim at quantitatively characterizing the corresponding avalanche statistics (i.e. size, duration, inter-avalanche time). Despite the coarsening dynamics is triggered by an internal driving mechanism, we find quantitative indications that such avalanche statistics displays scaling-laws very similar to those observed in the response of disordered materials to external loads.
\end{abstract}


\maketitle

\section{Introduction}

\begin{figure*}[t!]
\begin{center}
\includegraphics[scale=0.5]{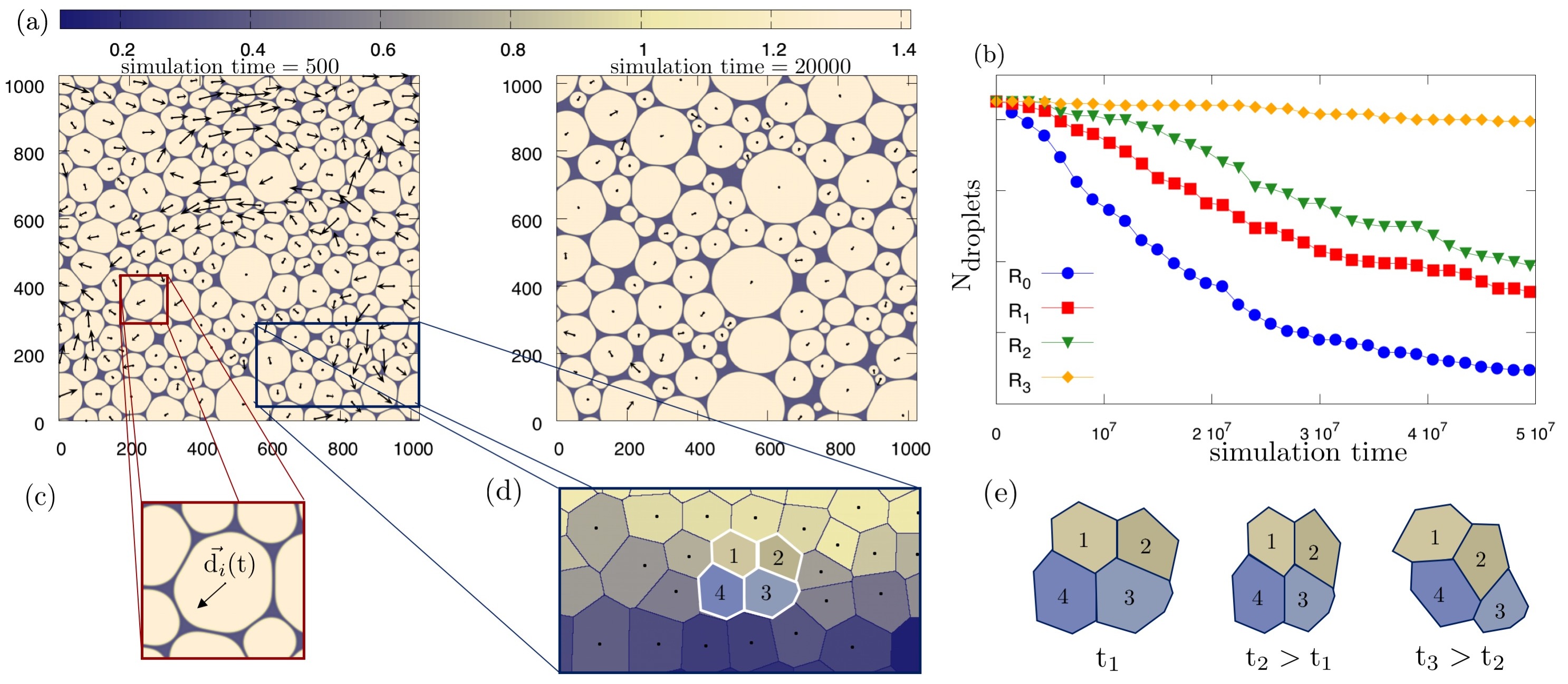}
\caption{Coarsening dynamics in a biphasic system consisting of closely packed ``droplets'' (light colors) dispersed in a continuous matrix (dark color) via lattice Boltzmann simulations~\cite{Benzietal14,Bernaschi16}. Panel (a): the system is prepared in an initial (slightly polydisperse) configuration and evolves through coarsening dynamics. Panel (b): number of droplets as a function of time for different simulations, corresponding to different coarsening rates, from the fastest ($R_0$) to the slowest ($R_3$). Panel (c): arrows indicate the displacement of droplets. Panel (d): we use the numerical tool~\cite{Bernaschi16} to build the Voronoi's tessellation of droplets and identify cells which undergo plastic rearrangements, highlighted by thick Voronoi edges. Panel (e): a plastic rearrangement is detected when a neighboring swapping occurs~\cite{CantatFOAMS}. \label{fig:coarsening}} 
\end{center}
\end{figure*}
Emulsions and foams are ubiquitous in nature and technology~\cite{Larson,Coussot}. These systems are {\it biphasic systems} with a complex microstructure: they are characterized by a collection of elementary constituents (i.e. liquid droplets or gas bubbles) dispersed in a continuous phase and stabilized against coalescence by the presence of surfactants at the interfaces. When the concentration of the dispersed phase is very large, the topological structure is that of soft domains with different sizes closely packed together and separated by thin films. The packing of the microstructure confers an elasticity to these systems~\cite{PrincenKiss89,Denkov09} and when driven with external loads they display an yielding behaviour with a rich intermittent dynamics~\cite{LulliBenziSbragaglia18}, typically characterized by a sequence of elastic loadings followed by their ``failure'' via abrupt {\it avalanche behaviours}~\cite{Barkhausen17,DurinZapperi00,Dastur81,Ruina83,Repainetal04,CantatPitois06,BonamySantucciPonson08,BarratReview17}. Such a kind of intermittent dynamics is observed not only for biphasic systems~\cite{CantatPitois06}, but also in many other contexts as diverse as the seismic motion of faults~\cite{Ruina83}, the propagation of cracks in fracture mechanics~\cite{BonamySantucciPonson08,LaursonSantucciZapperi10} and Barkhausen noise in ferromagnetic systems~\cite{Barkhausen17,DurinZapperi00}, just to cite a few examples.\\ 
From the micromechanical point of view, the avalanche behaviour hinges on the presence of plastic rearrangements~\cite{BarratReview17}. Plastic rearrangements can be viewed as topological changes/readjustments in the microstructure of the material; the effect of such readjustments propagates through the microstructure via the elastic interactions and can trigger novel rearrangements elsewhere, thus causing the avalanche behaviour. Avalanches are routinely characterized via the statistical description of their spatio-temporal properties, i.e. their size $S$~\cite{Linetal14,Sastryetal17,Salernoetal12,SalernoRobbins13,OkuzonoKawasaki95,Durian97,DenninKnobler97,KarimiFerreroBarrat17}, their duration in time $T$~\cite{GittingsDurian08,BudrikisZapperi13,LiuFerreroPuosi16}, and the inter-avalanche time $\tau$~\cite{GittingsDurian08,PerBak02,LaursonIllaAlava08,Sessoms10,Janicevicetal16}. In the recent years there has been an intense scrutiny to quantitatively understand whether the avalanche behaviour of different systems could be classified into different universality classes \cite{Salernoetal12,SalernoRobbins13,BudrikisZapperi13,Linetal14,LiuFerreroPuosi16}. In the majority of the studied cases, the systems are driven with an {\it external driving} and typically in a quasi-static protocol, i.e. in the limiting case of an infinitesimally small external load. It has to be noted, however, that the heterogeneity in the microstructure peculiar of biphasic systems may cause the system to be in non equilibrium states even in the absence of external loads. Equilibrium is then attained via the {\it coarsening dynamics}~\cite{Cugliandolo}. For foams, for example, coarsening materializes via the diffusion of the dispersed phase through the thin interfaces~\cite{Stavans93,vonNeumann,Lambert10,Webster01,SaintJalmes02,VeraDurian02,Mullins85,CantatFOAMS}:  domains slowly change in size, eventually creating local topological changes and plastic rearrangements, whose effect is propagated through the elastic microstructure~\cite{GittingsDurian08}. This poses the question of whether the coarsening dynamics is able to trigger an avalanche behaviour or not and its similarity with respect to the avalanche behaviour observedfor externally driven systems~\cite{DenninKnobler97,Dennin04}. The present communication aims to address this question.\\

\section{Numerical method}

The results shown in this paper are obtained via numerical simulations based on the lattice Boltzmann method developed by some of the authors in the recent years~\cite{Succi18}. In a nutshell, the method allows the numerical simulation of closely packed droplets of a dispersed phase in another continuous phase. Droplets are separated by diffuse interfaces where mesoscale interactions are ad-hoc tuned to prevent droplets coalescence. In Fig.~\ref{fig:coarsening} we report some relevant information on the coarsening dynamics that we can simulate using the numerical simulations. We analyzed a two dimensional biphasic system of droplets prepared in a slightly polydisperse configuration with a large packing fraction (see Fig.~\ref{fig:coarsening} (a)). The boundary conditions applied are periodic in both directions. The system is left free to evolve in time via the internal coarsening dynamics, which causes domains to move without any preferential direction in space (see Fig.~\ref{fig:coarsening} (c)). The numerical tool we used permits to compute the vectorial displacement of all droplets $\vec{d}_i(t)$ ($i=1...N_{\mbox{\tiny droplets}}$) at any time step of simulation; additionally, based on the procedures described in~\cite{Bernaschi16} we can quantitatively analyze topological changes in the droplets configurations via Voronoi Diagrams (see Fig.~\ref{fig:coarsening} (d)). By performing a Voronoi tessellation of the centers of mass of the droplets we can monitor the droplets neighbours and detect a ``rearrangement'' when these neighbours change. This allows to look at the location of rearrangements in space and time, as well as the size of the droplets involved in a rearrangement. Another key asset of the numerical simulations is the possibility to tune the mesoscale interactions at the interface to reduce or enhance the diffusion of mass, thus resulting in different coarsening rates. In Fig.~\ref{fig:coarsening} (b) we show results for different simulations corresponding to four different coarsening rates: from $R_0$ (the fastest) to $R_3$ (the slowest) \footnote{The different coarsening rates have been obtained by using the model described in~\cite{Benzietal15} by changing the nearest neighbors and next-to-nearest neighbors interactions. Specifically: ${\cal G}_{AA,1}={\cal G}_{AA,2}=-7.0$ and ${\cal G}_{BB,1}={\cal G}_{BB,2}=6.0$ ($R_0$); ${\cal G}_{AA,1}={\cal G}_{AA,2}=-7.3$ and ${\cal G}_{BB,1}={\cal G}_{BB,2}=6.4$ ($R_1$); ${\cal G}_{AA,1}={\cal G}_{AA,2}=-7.4$ and ${\cal G}_{BB,1}={\cal G}_{BB,2}=6.5$ ($R_2$); ${\cal G}_{AA,1}={\cal G}_{AA,2}=-7.45$ and ${\cal G}_{BB,1}={\cal G}_{BB,2}=6.55$ ($R_3$).}.  This is particularly important for the aim of the present paper, since this allows for a quantitative characterization of the avalanche statistics in a ``quasi-static'' protocol~\cite{Salernoetal12,SalernoRobbins13}, i.e. in the limit of vanishing coarsening rate. Further technical details are reported in~\cite{Benzietal14,Bernaschi16}. All dimensional numerical results are given in lattice Boltzmann units (lbu), not indicated hereafter for the sake of simplicity.\\ 

\begin{figure}[t!]
\begin{center}
\includegraphics[scale=0.17]{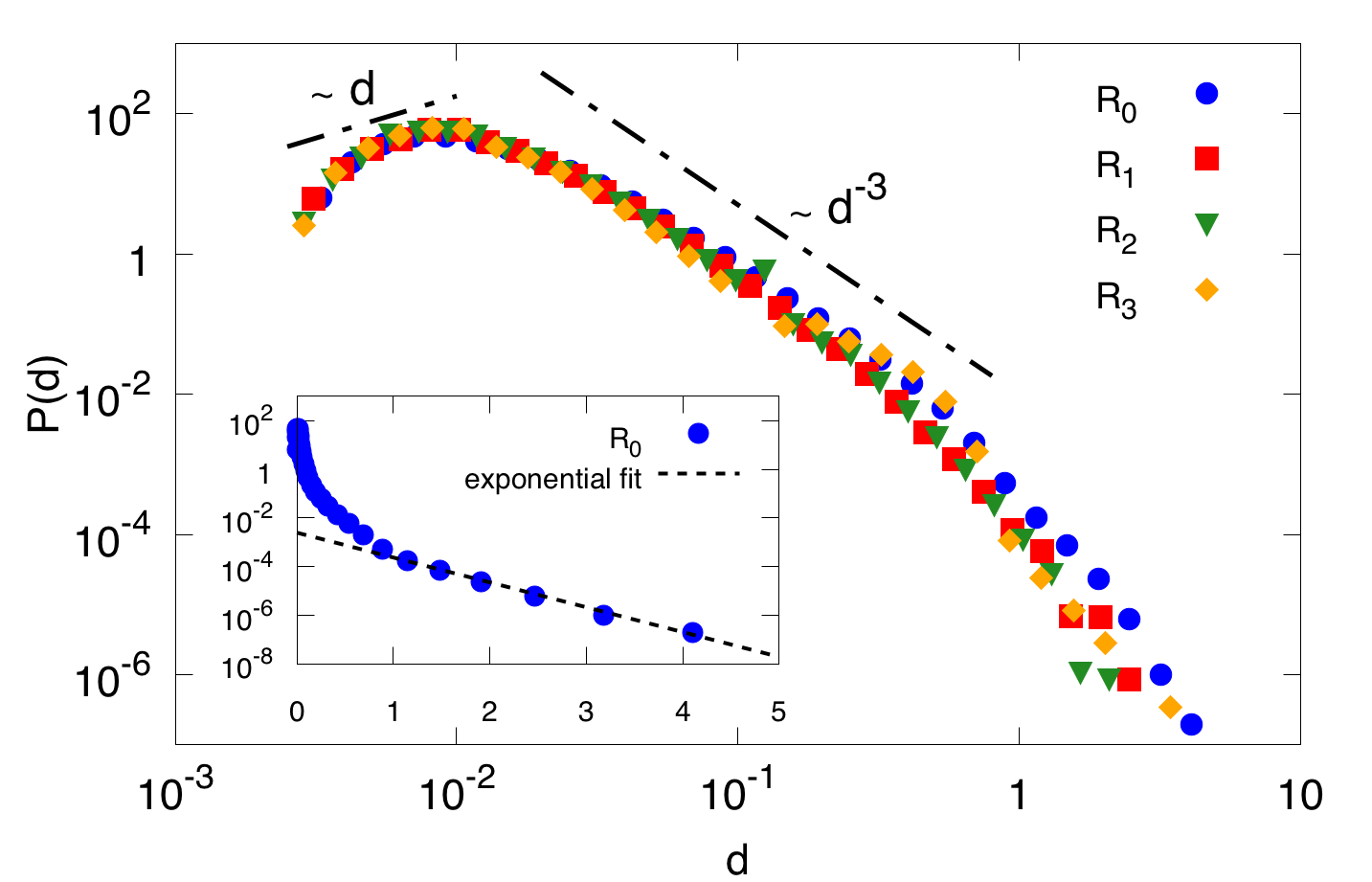}
\caption{Probability distribution function of the norm of the displacement field $d$ for different coarsening rates (cfr. Fig.~\ref{fig:coarsening}). The probability is computed by averaging over time and over the different droplets (see text for more details). We also report the scaling-laws for small ($P(d) \sim d$) and large ($P(d) \sim d^{-3}$) displacement norm that one can predict based on a mean-field argument for the ``elastic response''~\cite{FerreroMartensBarrat14,Chandrasekhar}. Inset: we show the log-lin plot of $P(d)$ to highlight the exponential tail in the distribution at large $d$. Data in the inset refer to the coarsening rate $R_0$ (cfr. Fig.~\ref{fig:coarsening}-(b)). \label{fig:chandrasekhar}}
\end{center}
\end{figure}
\begin{figure*}[t]
\begin{center}
\includegraphics[scale=0.5]{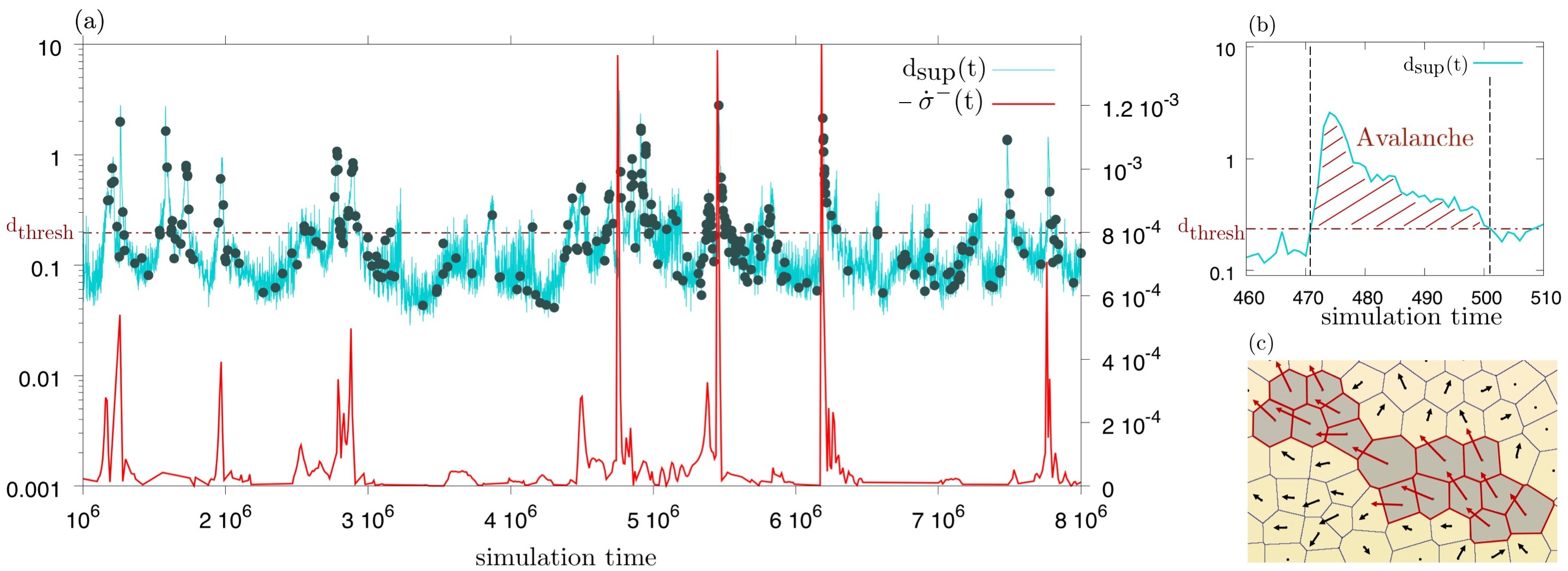}
\caption{Panel (a): time dynamics of the superior value of the displacement field $d_{\mbox{\tiny sup}}(t)=\mbox{sup}_{i} |\vec{d}_i(t)|$ (y-tics on left). Superimposed we report the evolution in time of minus the filtered (see text for details) stress derivative $-\dot{\sigma}^{-}(t)$ (y-tics on right). The occurrence of plastic rearrangements (cfr. Fig.~\ref{fig:coarsening}-(e)) is singled out with green bullets. Data refer to the coarsening rate $R_0$ (cfr. Fig.~\ref{fig:coarsening}-(b)). Panel (b): Protocol chosen to define an avalanche. A suitable threshold $d_{\mbox{\tiny thresh}}$ (red dashed-dotted line) on $d_{\mbox{\tiny sup}}(t)$ is introduced to select only extreme events. Panel (c): Protocol chosen to define the size $S$ of an avalanche based on $d_{\mbox{\tiny sup}}$: during the interval time in which $d_{\mbox{\tiny sup}}(t)>d_{\mbox{\tiny thresh}}$ we selected only the droplets with displacement that overcome the threshold $d_{\mbox{\tiny thresh}}$ (dark areas). Areas of selected droplets are summed upon to define the size $S$ of the avalanche. Notice that avalanches defined using this protocol do not necessarily correspond to connected domains in space.\label{fig:stressDisplPlastic}}
\end{center}
\end{figure*}
\begin{figure}[b!]
\begin{center}
\includegraphics[scale=0.15]{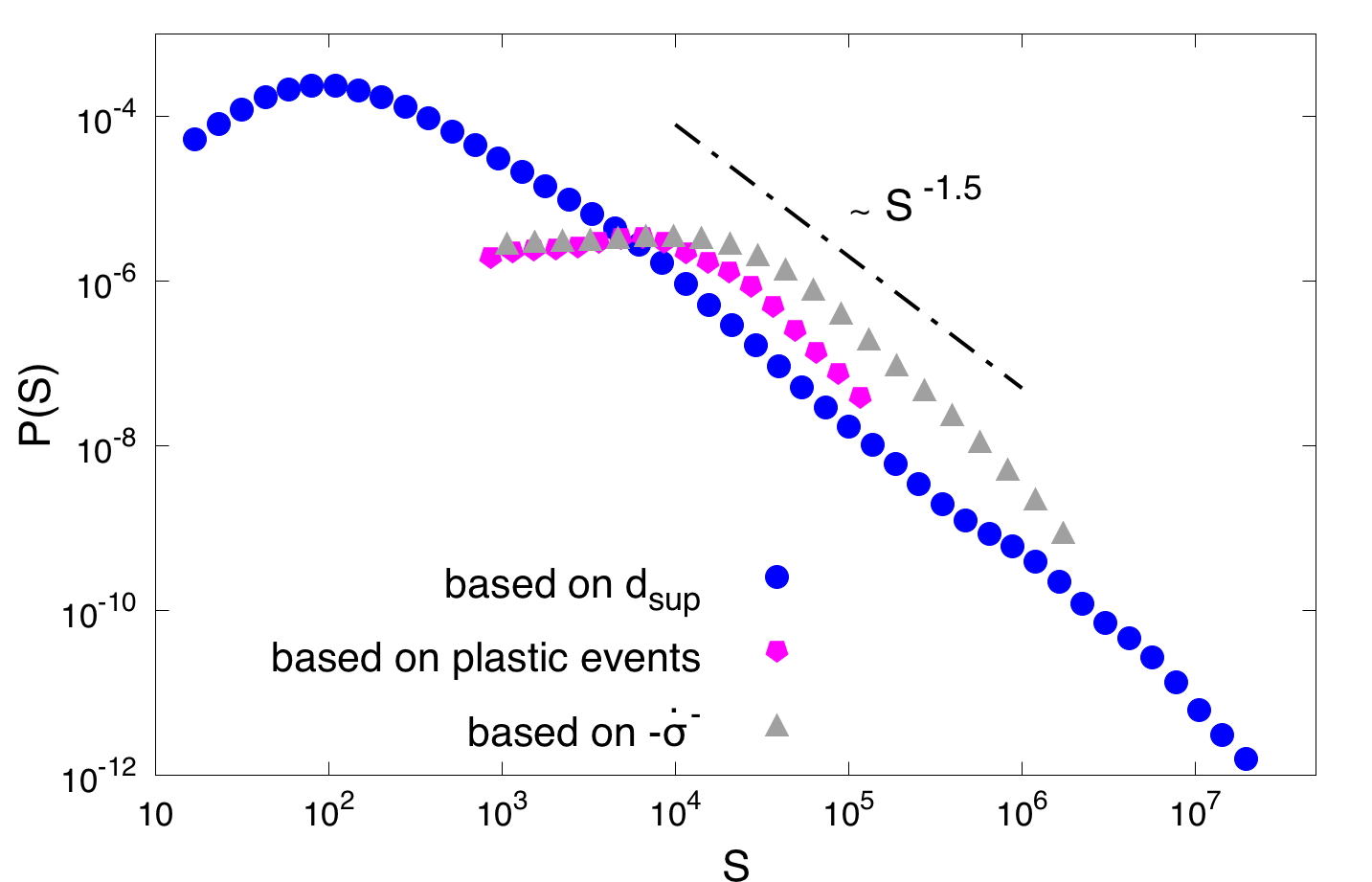}
\caption{Statistics of avalanche size. We report the probability distribution function (PDF) for the avalanche size $S$. The size $S$ is defined based on three different protocols: based on $d_{\mbox{\tiny sup}}$ (blue circles), based on $-\dot{\sigma}^{-}$ (grey triangles), based on plastic rearrangements (magenta pentagons). Details on the protocols are given in the text. Data refer to the simulation with the fastest coarsening rate $R_0$ (crf. Fig.~\ref{fig:coarsening}-b). We also report the mean-field scaling prediction~\cite{Sastryetal17,LiuFerreroPuosi16} (dashed line). \label{fig:sizeAvalancheT1}}
\end{center}
\end{figure}
\begin{figure}[t!]
\begin{center}
\includegraphics[scale=0.35]{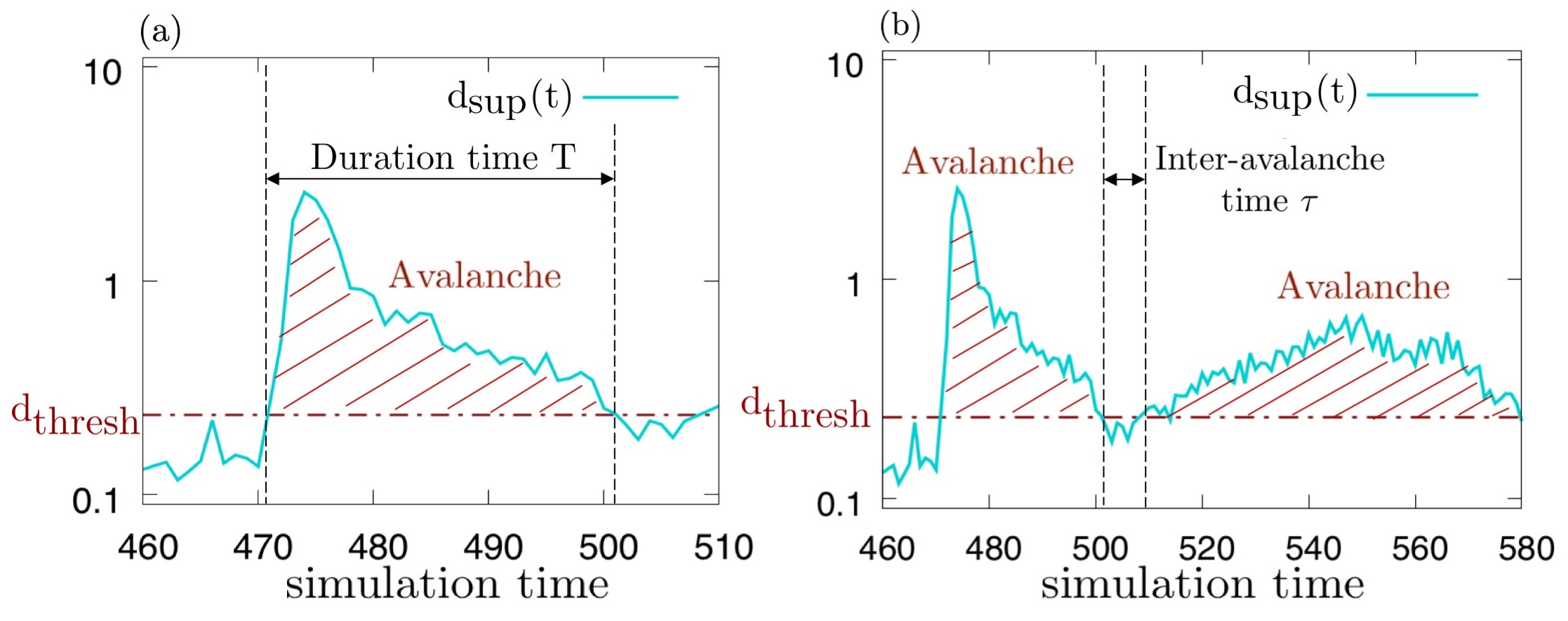}
\caption{Definition of the avalanche duration $T$ and the inter-avalanche time $\tau$. Panel (a): the avalanche duration $T$ is defined as the time interval in which $d_{\mbox{\tiny sup}}(t)>d_{\mbox{\tiny thresh}}$ (cfr. Fig.~\ref{fig:stressDisplPlastic}); Panel (b): the time lapse separating two avalanches defined as in Panel (a) provides a measure of the inter-avalanche time $\tau$. \label{fig:sketch}}
\end{center}
\end{figure}
\begin{figure}[t!]
\begin{center}
\includegraphics[scale=0.15]{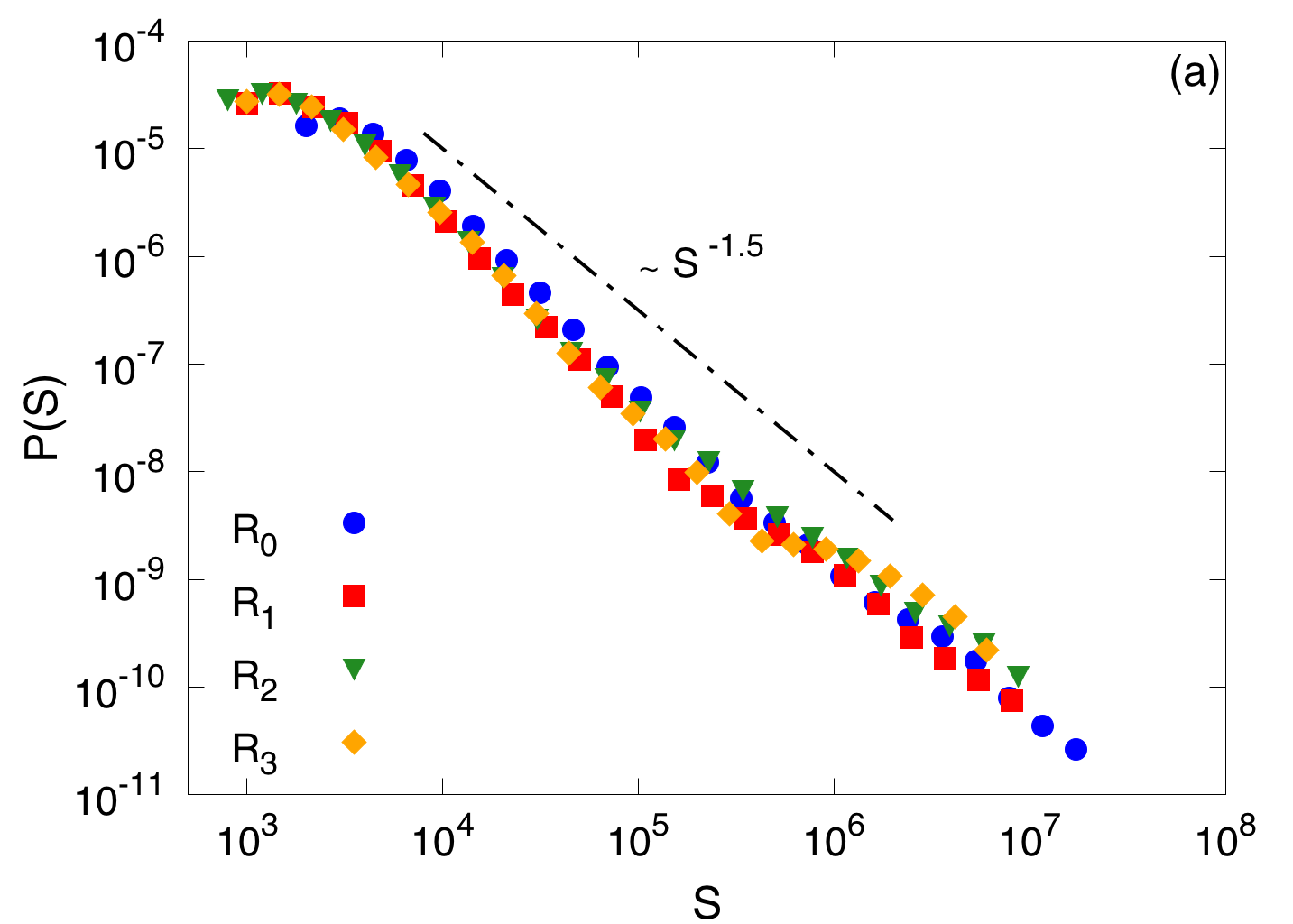}\\
\includegraphics[scale=0.15]{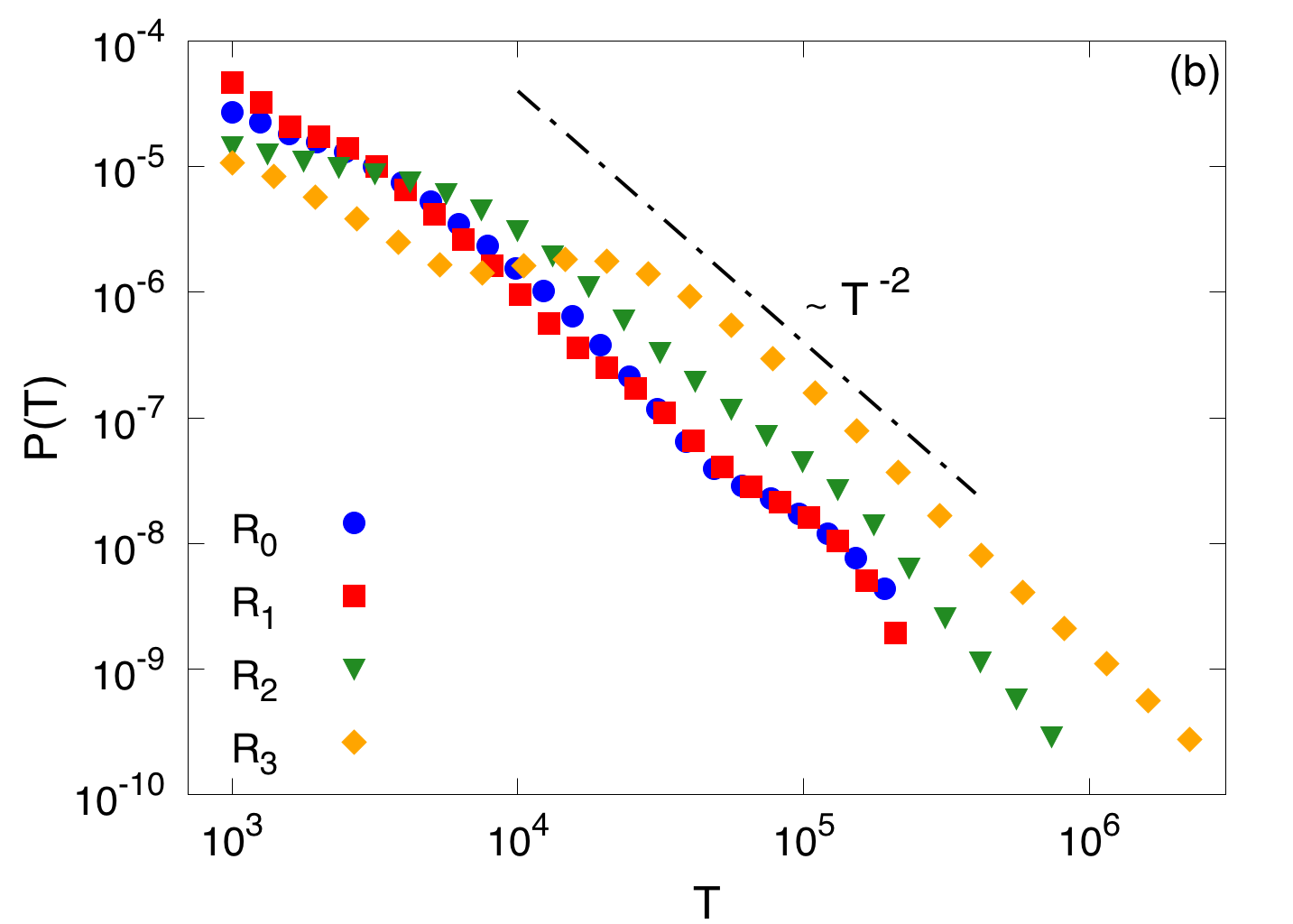}\\
\includegraphics[scale=0.15]{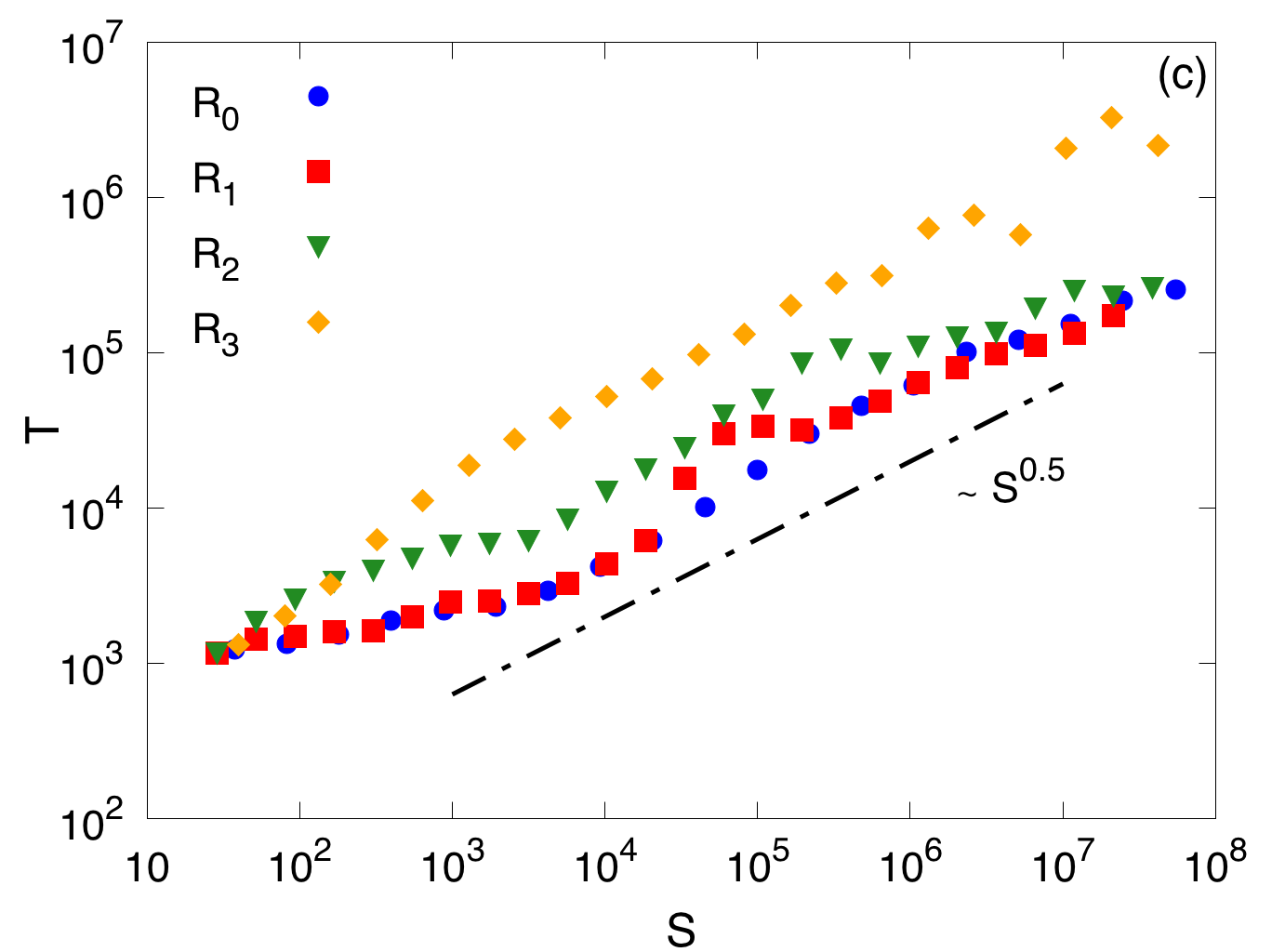}
\caption{Avalanche statistics for different coarsening rates. Panel (a): probability distribution function of size $S$. Panel (b): probability distribution function of duration time $T$. Panel (c): scatter plot of $S$ and $T$. For both $S$ and $T$, mean-field scaling predictions are also shown. \label{fig:pdfSizeDurationTime}}
\end{center}
\end{figure}
\begin{figure}[t!]
\begin{center}
\includegraphics[scale=0.15]{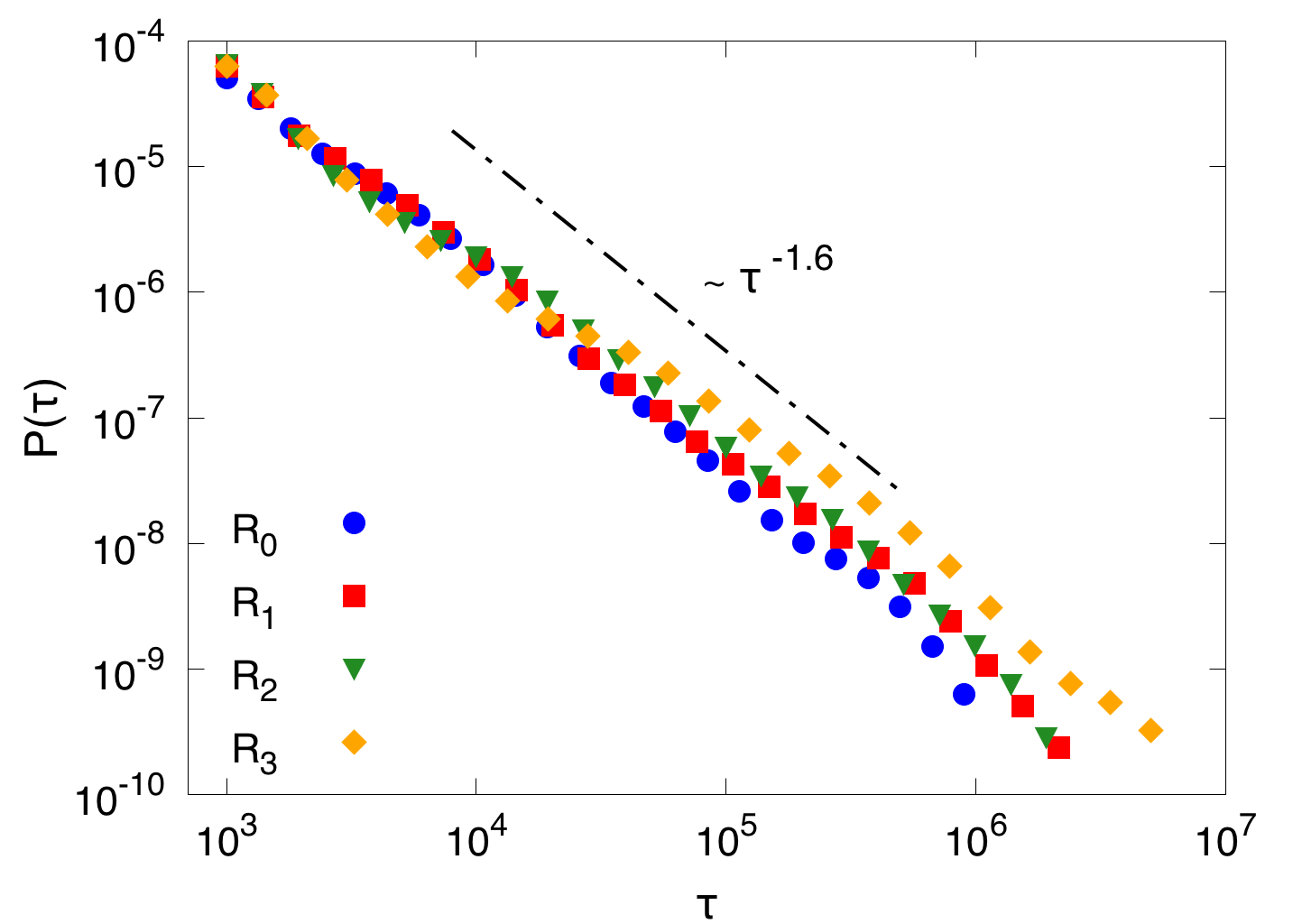}
\caption{Probability distribution function of inter-avalanche time $\tau$ for different coarsening rates. \label{fig:pdfTau}}
\end{center}
\end{figure}
\begin{figure}[h]
\begin{center}
\includegraphics[scale=0.15]{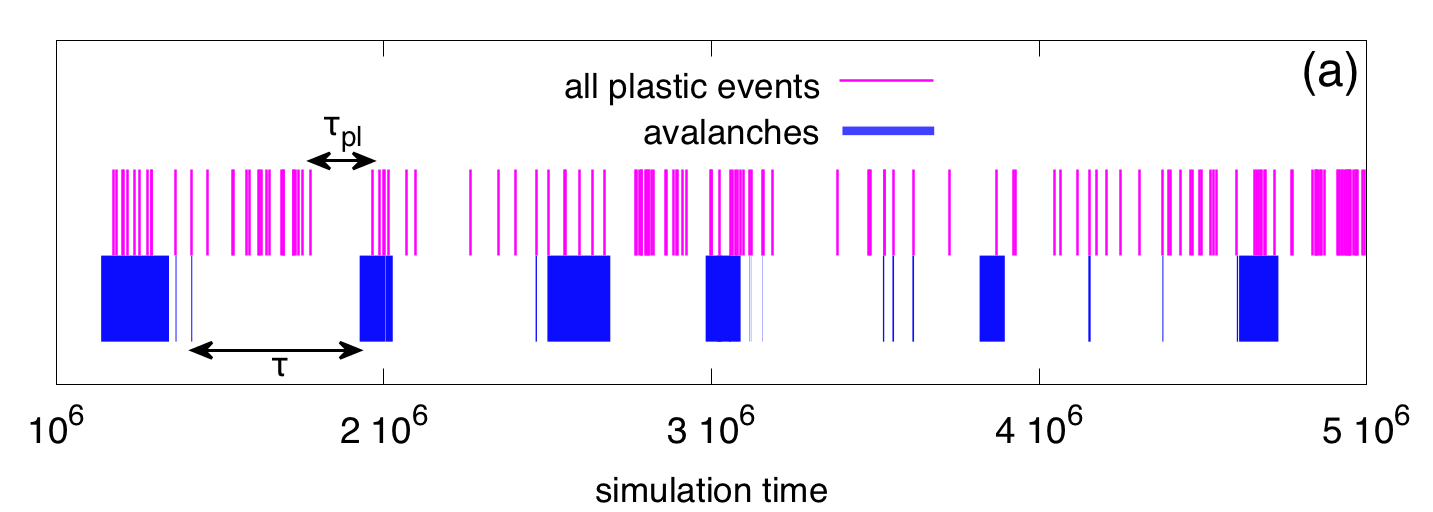}\\
\includegraphics[scale=0.15]{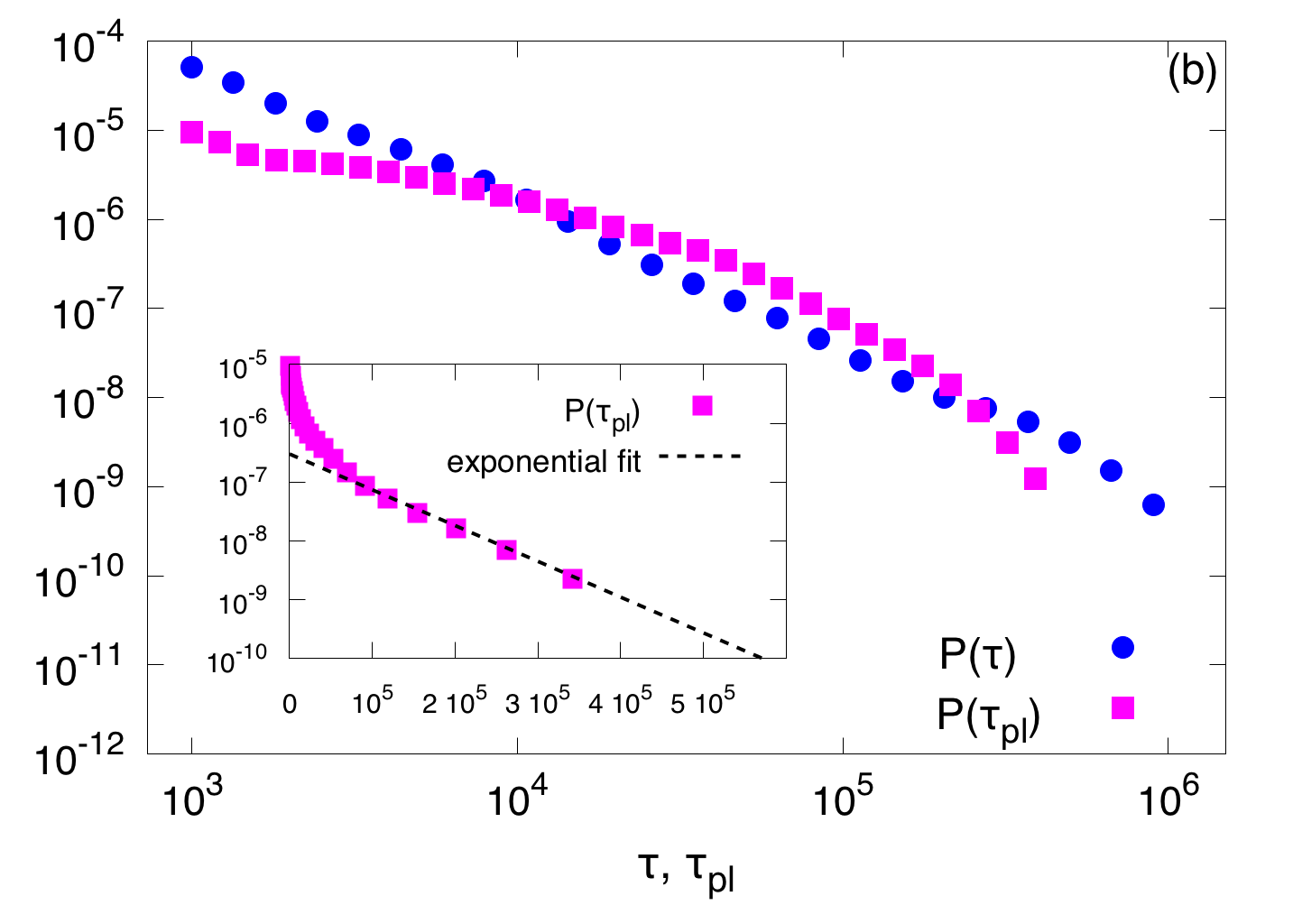}
\caption{The inter-avalanche time $\tau$ (cfr. Fig.~\ref{fig:pdfTau}) is compared with the waiting times between plastic rearrangements $\tau_{\mbox{\tiny pl}}$. Panel (a): for a selected time interval we report the occurrence of plastic rearrangements~\cite{ChristensenOlami92}. The upper sequence (magenta bars) refers to all plastic rearrangements detected in the system; the lower sequence (blue bars) only reports avalanches determined based on $d_{\mbox{\tiny sup}}(t)$ (cfr. Fig. \ref{fig:sketch}). Panel (b): we quantitatively compare the PDF of $\tau$ with the PDF of $\tau_{\mbox{\tiny pl}}$. While the PD of $\tau$ shows a clear power-law distribution, the PDF of $\tau_{\mbox{\tiny pl}}$ shows the power-law behaviour only at very short times, while exhibiting an exponential decay at larger $\tau_{\mbox{\tiny pl}}$ (see inset). Data refer to the simulation with larger coarsening rate $R_0$ (cfr.Fig.~\ref{fig:coarsening}-(d)).}\label{fig:interAvalancheTimePlastic}
\end{center}
\end{figure}

\section{Results and discussion}

As remarked before, during coarsening dynamics, the system may abruptly rearrange  (see Fig.~\ref{fig:coarsening} (e) for a sketch). Here we want to address the behaviour of collections of rearrangements (avalanches) mediated by the long-range elastic interactions. Hence it is crucial to have a quantitative assessment of the elastic response of the system. Following~\cite{FerreroMartensBarrat14,Chandrasekhar}, we estimated the probability distribution function (PDF) $P(d)$ of the norm of the displacement field $d_i(t)=|\vec{d}_i(t)|$, for all the time steps of the simulation ($t$) and all droplets ($i=1...N_{\mbox{\tiny droplets}}$). In presence of a long-range elastic response~\cite{FerreroMartensBarrat14,Chandrasekhar} one can predict some scaling-laws for $P(d)$ based on a mean-field argument, i.e. valid in the limit where the rate of events is small; one can then use the analytical result as a reference benchmark to assess the elastic response in our system. In such limit, $P(d)$ is expected to decay at large $d$ as $P(d) \sim d^{-3}$ for a two-dimensional system and $P(d) \sim d^{-5/2}$ for a three dimensional system; at small displacements, $P(d)$ is expected to behave as $P(d) \sim d$ for two dimensional systems and $P(d) \sim d^{2}$ for three dimensional systems~\cite{FerreroMartensBarrat14,Chandrasekhar}. These observations have already been documented in the literature, see for example studies on amorphous/soft solids in three dimensions~\cite{Sastryetal17,Bouzid17}, coarsening foams in three dimensions~\cite{GittingsDurian08} or elastoplastic models in two dimensions~\cite{FerreroMartensBarrat14}. In Fig.~\ref{fig:chandrasekhar} we report $P(d)$ for the various coarsening rates analyzed. While a clear scaling-law emerges for moderately large $d$, the small-$d$ asymptotic behaviour is barely visible, which we attribute to the fact that such small $d$ are well below the numerical lattice size and one would need to run simulations with larger droplets size to better highlight such scaling-law. We also notice that the crossover between the small-$d$ and large-$d$ behaviour is expected to depend on the rate of events~\cite{FerreroMartensBarrat14}, but we do not observe such dependence. This happens because, for the coarsening rates analyzed, the rate of plastic rearrangements is small; if we superimpose an external shear to the coarsening dynamics (data not shown), we observed that the two asymptotics at small and large $d$ change and the peak of the distribution of $P(d)$ shifts forward, in agreement with the results of~\cite{FerreroMartensBarrat14}. For very large $d$ the power-law behaviour exhibits an exponential cut-off (see inset) in agreement with~\cite{Sastryetal17}. Taken all together, the results displayed in Fig.~\ref{fig:chandrasekhar} provide a quantitative evidence of the elastic response of the system to plastic perturbations, and pave the way for the quantitative characterization of avalanches.\\ 
The characterization of the avalanches is routinely performed by monitoring the stress drop in the system~\cite{Sastryetal17,LiuFerreroPuosi16,Salernoetal12,SalernoRobbins13,Dennin04,Fretteetal96}. Key quantities are measured in relation to the released energy after an elastic loading, wherein a stress drop is observed in conjunction with plastic rearrangements that drive the system into a more stable configuration~\cite{Sastryetal17,OkuzonoKawasaki95,Durian97,DenninKnobler97}. The stress $\sigma$ is measurable in our numerical simulations~\cite{DolletScagliariniSbragaglia15} and in principle we could directly characterize the avalanches via the statistics of minus the filtered stress derivative, $-\dot{\sigma}^{-}$, computed by analyzing the evolution of the stress derivative $\dot{\sigma}(t)$ and selecting only the negative values (superscript $^{-}$). Nevertheless, being the elastic stress and the elastic response inherently related to the droplets displacement, it comes also natural to look for some definition of avalanches via the displacement field. The displacement field is expected to be very small in absence of plastic rearrangements; conversely, when plastic rearrangements occur, the displacement field is expected to strongly increase and localize, both in time and space. In other words, plastic rearrangements are expected to be responsible for the largest values attained by the displacement field, hence it is logical to look for the maximum displacement $d_{\mbox{\tiny sup}}(t)=\mbox{sup}_{i} |\vec{d}_i(t)|$~\cite{Benzietal16,LulliBenziSbragaglia18} as a non trivial quantity to address the statistical properties of avalanches. This was already highlighted in~\cite{Benzietal16}, where the very same model that we considered here was analyzed under the effect of an external driving in a Couette cell below yield. A stick-slip behaviour was observed with stress loads and sudden drops during avalanches. The maximum displacement was used to estimate the distribution of the energy release, displaying a power-law behaviour akin to the Gutenberg-Richter law of seismology~\cite{GR44}. The time evolutions of both $d_{\mbox{\tiny sup}}(t)$ and $-\dot{\sigma}^{-}(t)$ are displayed in Fig.~\ref{fig:stressDisplPlastic} (a). We also tracked in time the evolution of plastic rearrangements, indicated with a green bullet in the figure. Overall, Fig.~\ref{fig:stressDisplPlastic} (a) confirms the expectations posited before: the time dynamics of $d_{\mbox{\tiny sup}}(t)$ is that of a typical intermittent signal, with long ``rest'' periods (small $d_{\mbox{\tiny sup}}(t)$) separated by sudden intermittent peaks. Peaks in $d_{\mbox{\tiny sup}}(t)$ are accompanied by the occurrence of plastic rearrangements, which show some non trivial clusterization in time. Moreover, in correspondence of large values of $d_{\mbox{\tiny sup}}(t)$, we typically observe large $-\dot{\sigma}^{-}(t)$. These are compelling evidences that there exists a strong connection between $d_{\mbox{\tiny sup}} (t)$, $-\dot{\sigma}^{-}(t)$ and the fulfillment of plastic rearrangements, as already stressed in previous works (see for example~\cite{Dennin04,Jiangetal99,Chenetal12,Fretteetal96}). In passing, we remark that during the ``rest'' periods (low values $d_{\mbox{\tiny sup}}(t)$) without rearrangements, the system is undergoing a slow coarsening dynamics in quantitative agreement with the celebrated Von Neumann's law~\cite{vonNeumann} (see \cite{Benzietal15}, results not shown here). Moreover, notice that plastic rearrangements do not necessarily clusterize when $d_{\mbox{\tiny sup}} (t)$ is large, an issue that we will further discuss in Fig.~\ref{fig:interAvalancheTimePlastic}. \\ 
With these state of affairs, we proceeded into a more quantitative characterization of the avalanches, and we started from the avalanche size $S$. We estimated the PDF of the avalanche size $S$ using three different protocols: {\it i)} we define $S$ to be proportional to $-\dot{\sigma}^{-}$~\cite{Sastryetal17,LiuFerreroPuosi16,Salernoetal12,SalernoRobbins13,Dennin04,Fretteetal96}; {\it ii)} we define $S$ based on $d_{\mbox{\tiny sup}}$, as sketched in Fig.~\ref{fig:stressDisplPlastic} (b): first, we introduced a threshold $d_{\mbox{\tiny thresh}}$ to select extreme events. The threshold value $d_{\mbox{\tiny thresh}}$ is chosen by looking at the PDF of $d_{\mbox{\tiny sup}}$ (not shown), which displays two asymptotics: an increase at small values of $d_{\mbox{\tiny sup}}$ and a power-law scaling for large values of $d_{\mbox{\tiny sup}}$. The value of $d_{\mbox{\tiny sup}}$ in correspondence of the crossover between the two asymptotics is chosen as the threshold $d_{\mbox{\tiny thresh}}$, the rationale being that working above $d_{\mbox{\tiny thresh}}$ we are granted the selection of the extreme events. During the period of time where $d_{\mbox{\tiny sup}}(t)>d_{\mbox{\tiny thresh}}$ we selected only those droplets whose absolute value of displacements is above $d_{\mbox{\tiny thresh}}$ (see Fig.~\ref{fig:stressDisplPlastic}-(c)). Consequently, we defined $S$ as the sum of the areas of those droplets for the whole period of time where $d_{\mbox{\tiny sup}}(t)>d_{\mbox{\tiny thresh}}$; {\it iii)} finally, we define $S$ based on plastic rearrangements: when $d_{\mbox{\tiny sup}}(t)>d_{\mbox{\tiny thresh}}$ we selected only those droplets having performed plastic rearrangements and summed their areas. Notice that protocols {\it ii)} and {\it iii)} are based on a direct measure of droplets areas. Results on the PDF of the avalanche size based on the three protocols are displayed in Fig.~\ref{fig:sizeAvalancheT1}. The values of $-\dot{\sigma}^{-}$ have been rescaled by a prefactor to allow the three quantities to vary in the same range, without affecting their scaling properties. When the three PDF's display a scaling-law behaviour, the scaling exponent comes out to be very close to the mean field prediction~\cite{Sastryetal17,LiuFerreroPuosi16}. The main difference in comparing the three protocols emerges in the width of the region where the PDF displays a scaling-law: while the protocols based on the plastic rearrangements and $-\dot{\sigma}^{-}$ display a scaling-law behaviour on 1 and 2 decades respectively, the protocol based on $d_{\mbox{\tiny sup}}$ displays the scaling-law behaviour on more than 5 decades. \\
Motivated by the quality of the statistics on the avalanche size obtained via the threshold $d_{\mbox{\tiny thresh}}$ on the dynamics of $d_{\mbox{\tiny sup}}(t)$, we naturally extended the results to other coarsening rates (cfr. Fig.~\ref{fig:coarsening}-(b)) and also looked at the statistical properties of other avalanche quantities, like the avalanche duration $T$ and the inter-avalanche time $\tau$. Both $T$ and $\tau$ can be intuitively defined as sketched in Fig.~\ref{fig:sketch}. Results on the PDF's of these avalanche quantities are reported in Fig.~\ref{fig:pdfSizeDurationTime} (a), (b) and ~\ref{fig:pdfTau}. We remark that these PDF's have been obtained by focusing our attention only on those avalanches whose $S$ and $T$ display a scaling-law if plotted one against the other (see Fig.~\ref{fig:pdfSizeDurationTime} (c))~\cite{Dahmen11,BudrikisZapperi13,Budrikisetal16,LiuFerreroPuosi16}. The PDF's obtained for $S$ and $T$ exhibit a scaling-law behaviour, with the exponents reasonably independent of the coarsening rates and close to the mean field predictions, i.e. $P(S) \sim S^{-\alpha}$ with exponent $\alpha=3/2$ and $P(T) \sim T^{-\beta}$ with $\beta=2$~\cite{Sastryetal17,LiuFerreroPuosi16}. Regarding the inter-avalanche time, there is evidence of scaling-laws $P(\tau) \sim \tau^{-\gamma}$, suggesting a temporal correlation between avalanches~\cite{PerBak02,Corral03,Corral04,Baroetal12,Baresetal18}. Concerning the exponent $\gamma$, there is a more pronounced dependency on the coarsening rate and the scaling exponent is around $-1.6$. Notice that a power-law scaling for all $\tau$ with an exponent close to what we observe would give a diverging average inter-avalanche time; hence, our findings are somehow indicative of the very long relaxation dynamics peculiar of the system under study. Moreover, the results on the PDF for $\tau$ stimulate some comparisons with earlier results on coarsening foams~\cite{GittingsDurian08,Sessoms10} where it has been reported an exponential distribution for the PDF of the waiting times between successive plastic rearrangements $\tau_{\mbox{\tiny pl}}$. To better understand this point, in Fig.~\ref{fig:interAvalancheTimePlastic} we show a side-by-side comparison for the PDF of the inter-avalanche time $\tau$ and the just defined $\tau_{\mbox{\tiny pl}}$. In Fig.~\ref{fig:interAvalancheTimePlastic} (a) we report two sequences in a selected time interval: while the top sequence refers to the occurrence of all plastic rearrangements, the bottom sequence is a ``filtered'' one, in that it reports the plastic rearrangement at the time $t$ only if $d_{\mbox{\tiny sup}}(t)>d_{\mbox{\tiny thresh}}$. Although at this very qualitative level, one can perceive that the sequence of all plastic rearrangements is somehow more ``randomized'' than the filtered sequence. Going at a more quantitative level, we compared the PDF of $\tau$ with that of $\tau_{\mbox{\tiny pl}}$. Results are reported in Fig.~\ref{fig:interAvalancheTimePlastic} (b). The observed PDF's are manifestly different, with the PDF of $\tau_{\mbox{\tiny pl}}$ exhibiting a more evident exponential cut-off~\cite{Benzietal15,GittingsDurian08,Sessoms10} at large $\tau_{\mbox{\tiny pl}}$ (see inset). The observed difference in the PDF's highlights the importance of both spatial displacements and plastic rearrangements in characterizing the avalanche statistics. Indeed, clusters of rearrangements do not necessarily correspond to large displacements (see Fig.~\ref{fig:stressDisplPlastic} (a)) and they need to be properly considered in conjunction with large displacements to observe a neat power-law. This result bears similarities with some earlier observations on the spatio-temporal clustering of real earthquakes~\cite{KaganJackson91,ChristensenOlami92,Corral04,Mulargiaetal17}, where it is not trivial to distinguish a main earthquakes from aftershocks.

\section{Conclusions}
Summarizing, based on mesoscale numerical simulations, we have investigated the statistical properties of the coarsening dynamics of an elastic biphasic system consisting of soft elementary constituents (droplets) of a dispersed phase packed together in another continuous phase. While being driven by the ``internal'' coarsening dynamics, the system develops bursts of intermittent activity (i.e. avalanches) inherently associated to droplets displacements and rearrangements. To quantitatively analyze such dynamics, we have defined the observables peculiar of the avalanche statistics starting from the droplets displacements and analyzed their statistical properties. Results show evidences of power-law scalings, with spatio-temporal correlations that echo those measured in the avalanche statistics of amorphous systems driven with ``external'' loads. Scaling exponents are close to mean-field predictions, although a more precise determination of their values would require further numerical studies.\\

\section*{Acknowledgements}
The authors wish to thank F. Bonaccorso and M. Bernaschi for computational support. FP wishes to thank M. Lulli for his introductory help in controlling the computational tools developed in~\cite{Bernaschi16}. MS acknowledges financial support from the project "Hydrodynamics of Soft-Glassy materials through microdevices'' (HYDROSOFT) financed by the University of Rome ``Tor Vergata'' (Bando ``Mission Sustainability'').\\

\bibliographystyle{rsc}
\bibliography{francesca}

\end{document}